# Novel insights on swelling and dehydration of Laponite RD

G. A. Valencia,[a,b] M. Djabourov,[b] F. Carn,[c] and P. J. A. Sobral[a]

a. Department of Food Engineering, Faculty of Animal Science and Food Engineering, University of São Paulo, Pirassununga, SP, Brazil.

b. Laboratoire de Physique Thermique, ESPCI-Paris, PSL Research University, 10 Rue Vauquelin, 75231 Paris Cedex 5, France.

c. Laboratoire Matière et Systèmes Complexes, UMR 7057, Université Denis Diderot-Paris 7, Case Courrier 7056, 10 Rue Alice Domon et Léonie Duquet, 75205 Paris, Cedex 13, France.

* Corresponding author: Madeleine.Djabourov@espci.fr ; madeleine.djabourov@orange.fr

## Abstract

This paper investigates water sorption and dehydration of a model synthetic clay (Laponite) and highlights the large differences with natural clays (montmorillonite, hectorite) belonging to smectite family of structure and composition. Measurements were done by combining for the first time thermogravimetric analysis and X-ray diffraction methods on Laponite powders stored in atmospheres with controlled relative humidity. Laponite exhibits a very good reproducibility of sorption diagram and appears as the most hygroscopic among smectites. The interlayer space increases from 11 to 20 Å when $r = n_{H2O}/n_{Na}$ increases from 2 to 22 in agreement with recent numeric simulations by Morrow et al. We show that the desorption process, for $2 \leq r \leq 22$, in isothermal conditions follow a first order kinetic identifying two regimes around $r^* = 7$ which is close to the coordination number of Na in concentrated NaCl solution. Activation energies and frequency factors are derived for the two regimes of water desorption and discussed in relation with stability of the cationic complex.

## 1. Introduction

Smectite clays (Sm) are layered mineral nanostructures consisting of a central sheet composed of $Li^+$, $Mg^{2+}$, $Al^{3+}$ or $Fe^{3+}$ cations in octahedral coordination to oxygen atoms or hydroxyl groups. This central layer is sandwiched between two silicate layers where the silica atoms are in tetrahedral coordination to oxygen atoms. The presence of charge unbalance leads to an overall negative charge in the central layer, which is compensated by labile counterions, mainly $Na^+$ and $Ca^{2+}$, located at the surface of the silicate layers.

The Sm group incorporates a variety of clays with the most common ones being montmorillonite (Mt) and hectorite (Ht)[1]

Laponite (Lap) Lap (hydrous sodium lithium magnesium silicate) is a synthetic clay, a crystalline layered silicate colloid with structure and composition closely resembling Sm. Lap nanoparticles have a disk-shape with a thickness of approximately 1 nm, and a diameter of approximately 25 nm, much smaller than natural clays (microns). Lap disks possess a net negative charge which is balanced by the positive charge of sodium ions. Lap is used to modify rheological properties of liquids, cosmetics, paints, and inks [2].

More recently these anisotropic nanoparticles were considered as building basic units of multilayered photonic crystals[3]. Lap and surface modified Lap can be used for adsorption of dyes[4], surfactants and biomolecules, to reinforce biopolymers based films [5] or as effective drug carrier [6].

Although the swelling of natural clays was broadly studied in the context of catalysis or environmental industries [7–9], few studies were devoted so far to Lap, which however presents the advantage over natural clays to have smaller particle size, lower structural polydispersity (dimension, surface charge) and better stability in water compared for instance to Mt.

This paper reports new results on Lap swelling under different relative humidities (RH) by combining for the first time thermogravimetric analysis (TGA), X ray diffraction and water desorption kinetics. We reveal the unusual characteristics of Lap RD upon dehydration and discuss the relation with the clay structure.

## 2. Experimental

The synthetic clay is Lap RD (Southerm Clay Products Inc.) with molecular composition $(Si_8[Mg_{5.5}Li_{0.4}H_{4.0}O_{24.0}]^{0.7-}[Na_{0.7}]^{0.7+})$[10]. Lap RD initially stored at room conditions was transferred to desiccators equipped with hygrometers containing silica gel with RH=0.1%, 3% and saturated salt solutions for RH values of 31%, 45%, 71%, 89% and 94%. Powders were equilibrated during 4 days under each RH value (powders reached a constant weight). The equilibrium water content (moisture content) $w_{equil}$ is given by:

$$w_{equil} = \frac{m_{water}}{m_{dry\ clay}} \qquad (1)$$

The mass of dry Lap RD is determined after heating the powders at 150°C using thermogravimetric analysis.

After equilibrium, Lap RD was analyzed using an X-ray diffractometer (Panalytical Empyrean, Cu Kα radiation) equipped with a multichannel detector PIXcel 2D solid-state detector (255 x 255 pixels, each pixel is 55 µm x 55 µm). X-ray diffraction patterns are in the 1.5-70° range, with a 0.025° step size and 30s per step. Patterns were analyzed using Panalytical HighScore Plus software. The $d_{001}$ value of basal reflection was calculated using Bragg law:

$$\lambda = 2d\sin\theta \tag{2}$$

where $\lambda$, is the wavelength of Cu K$\alpha$ radiation (1.54187 Å), $d$ is the spacing and $\theta$, is the reflection angle. The overall duration of a measurement is about 10 min.

Water desorption rates of Lap RD were investigated by TGA (Q5000, TA Instruments, USA) in isothermal and non-isothermal modes operating with constant pure nitrogen gas flow (25mL/min) and with open Pt pan with 10mm diameter [11,12]. The initial sample weight was approximately 50mg. The dehydration rates are analyzed in isothermal conditions at 35°C and 45°C during 120min, heating rate +1°C/min, from 20°C to 150°C, continued by 60min isothermal drying at 150°C. Experiments were repeated at least two times for each sample and showed a good reproducibility.

## 3. Results and discussion

### 3.1 Equilibrium water uptake

Lap RD exhibits a type II isotherm (Fig. 1a), typical for the adsorption of gases on solid surfaces [13]. Our results Fig. 1a are in good agreement to those reported by [14] at a temperature of 31°C.

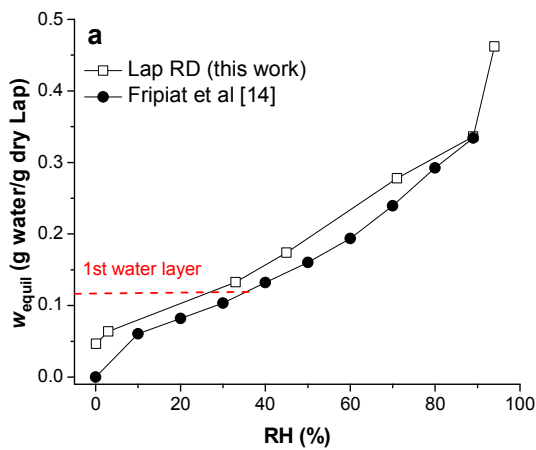

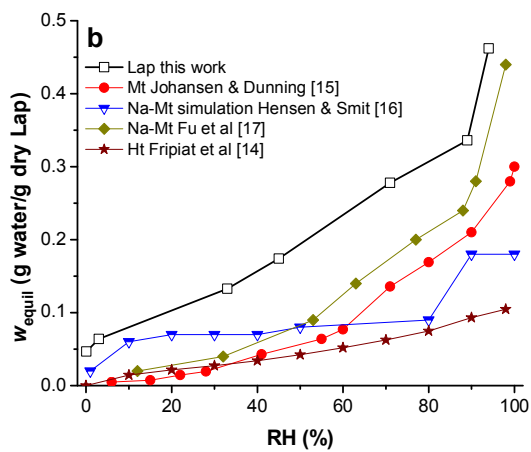

**Fig. 1a.** Sorption isotherm of Lap RD (squares) conditioned under different relative humidities at 23°C and results from [14]; **b.** Comparison between the sorption isotherms of Lap RD (this paper) and other Sm clays: (Mt) [15] full disc symbols; simulations [16] semi-solid triangle symbols; [17] full diamond symbols; (Ht) [14] star symbols.

Fig. 1b shows the comparison between water sorption of Lap RD and of other natural clays. Lap RD appears much more hygroscopic than other clays on the whole RH range and the measurements performed on this clay appear much more reproducible. These features suggest that the equilibrium adsorption may occur only with Lap, which provides the best model system for testing the swelling mechanisms.

Interpretation by [14] of the adsorption isotherm of Lap locates the first hydration layer around RH≈20% with $w_{equil}$ ≈0.11 g/g dry Lap, whereas in Mt clays the 1rst layer forms RH≈60 %[18]. According to [13] the specific surface area of Lap RD to water is 345.4m$^2$/g, being at least three times larger than in natural clays. Large specific surface areas can be interpreted as high proportion of active sites for water adsorption. However, the estimated specific surface area of phyllosilicates from crystallographic determinations is 750m$^2$/g [14,19].

In order to elucidate the water adsorption mechanisms in Lap RD crystals, X-ray diffraction measurements were performed on the humid clays after equilibration in desiccators.

## 3.2 Structure evolution as a function of interlayer hydration

Distinct visible reflections appear in X-ray diffractograms (Fig. 2) at 3.7°≤2θ≤7.7° ($d_{(001)}$) and 2θ=60.8° ($d_{(060)}$) which are related to the interlayer space and to the trioctahedral character of Lap, respectively [20–22]. Reflections with 2θ<10° are visibly shifted to the left with increasing humidity. Swelling of the clay modifies the distance $d_{(001)}$ related to the interlayer space. The $d_{(001)}$ values calculated from the position of the low angle diffraction peak increase linearly with $w_{equil}$, (derived from Fig. 1a) following the equation:

$$d_{(001)}(\text{Å}) = 21.86 * w_{equil} + 10.21 \quad (3)$$

suggesting that water molecules adsorb in interlayer space and produce a physical swelling. The distance extrapolated towards $w_{equil}$=0 is $d_{(001)}$=10.2 ± 0.2Å and predicts the interlayer space when Lap RD is completely dehydrated, very close to $d_{(001)}$=10.3Å reported by [16] from numerical predictions in dehydrated structures of Mt.

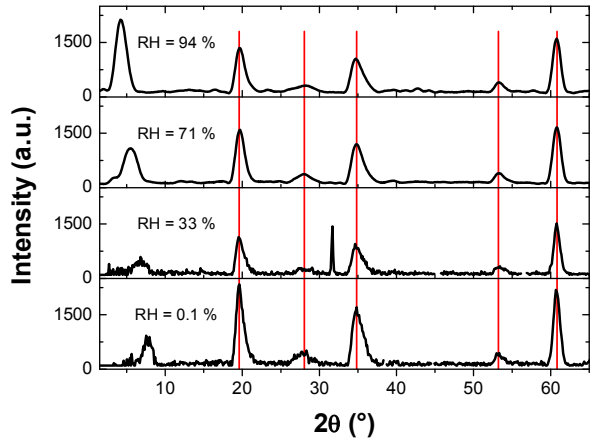

**Fig. 2.** X-ray diffraction patterns of Lap RD conditioned under different relative humidities with 1.5°<2θ<70°. The peak at the smaller scattering angle is related to the interlayer space which increases with water adsorption.

Recent numerical simulations on St type clays follow two different strategies in interpreting the swelling process: [16] predicts $w_{equil}$ versus RH for Na-Mt, as shown in Fig.1b and [23] predicts $d_{(001)}$ versus the ratio $r$=H$_2$O/Na$^+$ for Na-Ht.

Molecular dynamics and Monte Carlo sampling simulations by [16] show that the crystalline swelling of Na-Mt is rigorously determined by the number of water molecules which enter the interlayers via the presence of the cation. Since water vapor adsorption and swelling are controlled by the cationic composition in Mt, is it interesting to calculate the ratio $r$=H$_2$O/Na$^+$ for Lap equilibrated in desiccators. With the chemical composition given by [10] the Lap RD unit cell contains 0.7Na$^+$ ions for a molar mass of 765.18g/mol. $w_{equil}$ allows calculation of the ratio $r$. The results are shown in Fig.3. The lowest RH exhibit the lowest $r$ value with $r$=2.2 and the largest value is $r$=22. Three main differences appear with natural clays: i) Hydration in Lap RD is very high compared to Na-Mt where this ratio is limited to the range $r$<10. ii) Simulations by Morrow at al 2013 [23] clearly predict a step like increase of the interlayer distance $d_{(001)}$ *versus parameter $r$* as shown in Fig.3. iii) Predictions for water adsorption $w_{equil}$ in Na–Mt by Hensen and Smit [16] *versus RH* reported on Fig. 1a, show large domains where $w_{equil}$ remains constant (plateau values), which means plateaus in the layer spacing versus RH (constant $w_{equil}$ means constant $d_{(001)}$ ).

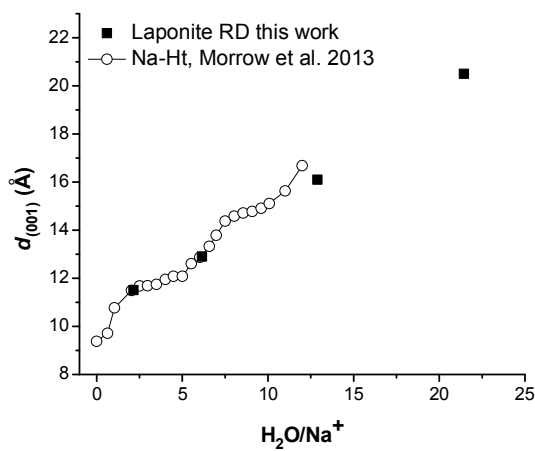

**Fig. 3.** Interlayer space (Å) versus hydration level ($r=H_2O/Na^+$) for Lap RD (this work) (squares) and results on Na-Ht from simulations by [23] (circles).

Our measurements of the interlayer spacing Fig.3 agree with the molecular dynamics simulations on Na-Ht between $r\approx 2$ and $r\approx 13$. Because Lap RD hydration is much more favorable, $r\approx 13$ is obtained for RH=71% with the interlayer $d_{(001)}$=16.1Å and the distance still increases to $d_{(001)}$=20Å with RH=94%. However, there is no evidence in our results of any plateau values either versus RH or versus $r$.

Hensen and Smit explain the mechanism of clay swelling and the interplay with the cation hydration: between $r$=3 and 4 water molecules are positioned in the center of the interlayer. The increase in layer spacing results in less confinement of the water molecules allowing them to orient to the sodium counterions. Hydration is facilitated by the increased interlayer volume. It appears notably that the sodium ions move to the interlayer center for the ratio $r$=6: this value corresponds to the sodium coordination number in concentrated NaCl aqueous solutions.

Other Monte Carlo simulations of Na-Mt by Meleshyn and Bunnenberg [24] found the "termination" of crystalline swelling at a layer space of $d\sim$19Å by formation chain like structures consisting of Na cations, water molecules, and oxygen. The authors suppose that such a persistent structure "locks" the interlayer space, until excess water is able to break this chain by osmotic forces. We did not reach the "termination" step and Lap remained crystalline at the highest RH.

The differences between Lap and natural clay swelling are therefore significant. One important difference is the size of the platelets which is much smaller in the synthetic clay (Lap) 25 nm than in any other Mt clays (micron size). The differences may explain the large discrepancies in the adsorption diagrams versus RH of natural clays, on one hand, and of natural clays and Lap on the other hand. The equilibrium adsorption may occur only with Lap, which provides the best model system for testing the swelling mechanisms.

Our analysis also explains the difference between the determinations of the specific surface area of Lap by crystallography (750m$^2$/g) and by BET theory (345m$^2$/g). The latter model determines therefore the adsorption of a monolayer of water molecules between a *pair of layers* (plus the small contribution from a monolayer on edges and external surfaces), in agreement with the model based on complex formation at equilibrium between interlayer cations and water molecules.

### 3.3 Thermogravimetric analysis of isothermal desorption

We carried TGA experiments at constant temperatures following precisely the mass loss versus time after samples were equilibrated in different humid atmospheres. The mass of water was calculated and normalized by the final mass of Lap RD measured after 120 min. $x$ (moisture content) is the normalized mass of water by the final mass of Lap RD:

$$x(t) = \frac{m(t) - m_{final}}{m_{final}} \tag{4}$$

where $m(t)$ is the sample mass (mg) at the time $t$ (min), and $m_{final}$ is the final mass (mg) of sample. Samples analyzed with these isothermal protocols have a small amount of residual humidity approximately 0.01g of water/g of dried Lap determined by heating at 150°C. It was considered that the "final mass" of the Lap after 120 min corresponds to "dry" Lap.

The normalized mass, $x$ versus time and the derivative of normalized mass, $|dx/dt|$ versus time are reported in Fig. 4a and 4b. The initial values of $x$ vary between 0.45 and 0.05 according to RH, then $x$ decreases versus time. The final value is reached after 120 min (Fig. 4a). When $t=0$min, the most humid samples exhibit large $|dx/dt|$ values (Fig. 4b) due to the large difference between the water vapor concentration at the surface of Lap and in (dry) nitrogen gas. A change of the slope of $|dx/dt|$ appears versus time (see inserted arrows in Fig. 4b) for Lap conditioned at RH=71%, 89% and 94% after 4.7min, 5.49min and 11.5min, respectively. The $x$ value associated to this particular moment is 0.15 ± 0.02g water/g dry Lap whatever the conditioning RH. In contrast, Lap conditioned in RH≤45% shows no such a change because initially $x$≤0.14g water/g dry Lap. There is a visible change in the drying rate associated with the passage from high to low water contents.

There is a formal analogy between dehydration and the desorption of gas from a solid surface or with decomposition reactions of solids(see for instance [25]). We assume here that desorption of water molecules adsorbed on Lap is a first order reaction and this assumption is validated with the experimental data.

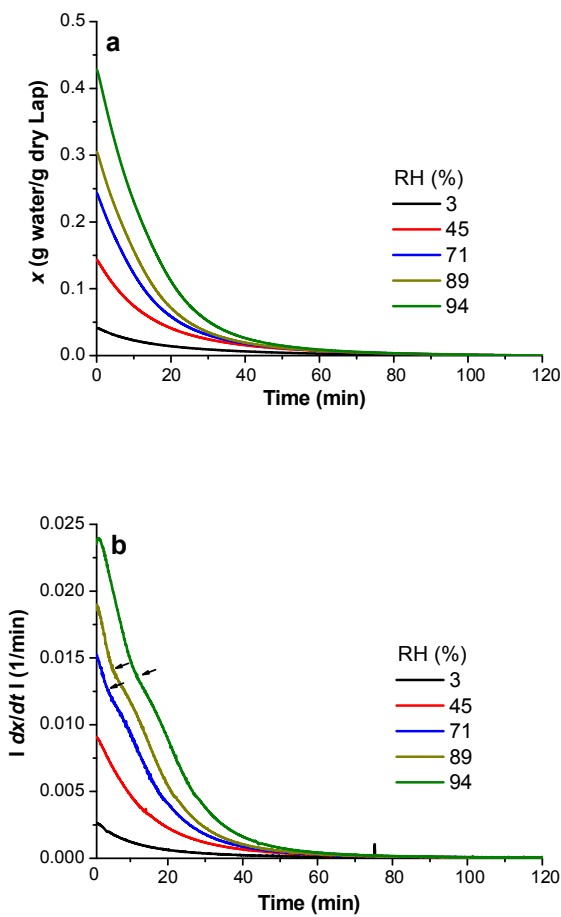

**Fig. 4a** Normalized mass $x$ versus time; **b.** Derivative of normalized mass, $|dx/dt|$ versus time of Lap conditioned under different relative humidities and analyzed in isothermal conditions at 35°C.

The variable $x$ is the fraction of occupied sites at the surface of the solid (or surface coverage), $dx/dt$ is the desorption rate. Assuming that there is no re-adsorption, the reaction rate for a first order reaction writes [25]:

$$-\frac{dx}{dt} = Axe^{-E/RT} \quad (5)$$

where $A$ is a frequency factor (1/min), $E$ is the activation energy (kJ/mol), $R$ is the universal constant gas (kJ/mol·K), and $T$ is the temperature (K).

From Eq. (5) is possible to anticipate that desorption kinetics in isothermal conditions can be expressed as a potential function, where $|dx/dt|$ decreases as a function of time, following the decrease of $x$. In isothermal conditions, activation energy can be calculated by means of Eq. (5); measuring the desorption rates at least at two constant temperatures allows determining both $A$ and $E$ values without any adjustable parameter.

The plots of $|dx/dt|$ versus $x$ derived from several experiments (RH=3%, 45% and 94%), dried in isothermal conditions, both at 35°C and 45°C are showed in Fig. 5a. Interestingly, in each plot, at a fixed temperature, all desorption rates $|dx/dt|$ versus $x$ are superposed on a single curve. As predicted by Eq. (5) there is a linear dependence of the slope $|dx/dt|$ versus $x$ at a fixed temperature. However, it is observed distinctly two regimes according to the water content of Lap: the slope changes when $x^* \approx 0.15 \pm 0.02$ g water/g dry Lap. Desorption rates are larger when temperature increases in both regimes, which is expected from Eq. (5). In the first regime, with $x<0.15$, the linear regression gives respectively the slopes 0.074 and 0.137 at 35° and 45°C, from where the activation energy is calculated. The second regime, starts at $x^* \approx 0.15$ and the slopes of the regression lines are 0.055 and 0.083 at 35° and 45°C. It is derived from these plots the following desorption energies and frequencies:

$$E_1 = 52 \pm 3 \text{ kJ/mol, when } x<0.15, \text{ and } A_1 = 1.4 \cdot 10^8 \pm 4 \cdot 10^7 \text{min}^{-1}$$

$$E_2 = 33 \pm 1.5 \text{ kJ/mol, when } 0.15<x\leq 0.4, \text{ and } A_2 = 4.6 \cdot 10^4 \pm 9 \cdot 10^2 \text{min}^{-1}$$

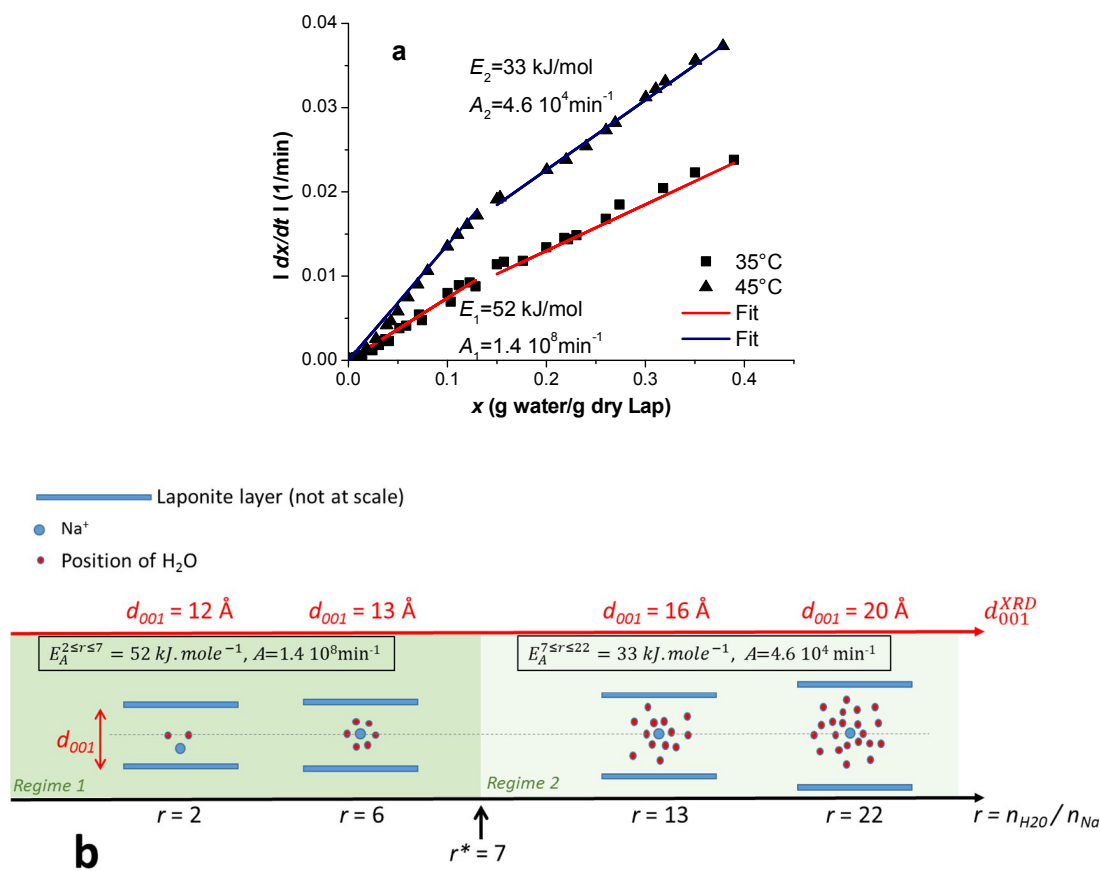

**Fig. 5a.** Derivative of normalized mass, $|dx/dt|$ versus normalized mass $x$ of Lap conditioned at RH=3%, 45% and 94%, and analyzed in isothermal conditions at 35°C and 45°C. Straight lines represent the fit by linear regressions; **b.** Schematic view of the interlayer structure of Lap upon

10swelling: the upper scale is the $d_{001}$ distance measured by X ray diffraction and the lower scale represents the hydration ratio *r*. In simulations, the water molecules are located near the $Na^+$ ion and progressively diffuse in the interlayer for large *r* values.

When $x = x^* \approx 0.15 \pm 0.02$g water/g dry Lap, $r^* = 7.13 \pm 0.47$. The particular value $r^*$ when the transition occurs between the two regimes, is close to that predicted by [16] for Na-Mt (*r*=6) corresponding to $Na^+$ coordination number of concentrated NaCl solutions. Fig.5b shows schematically the relation between the distance $d_{001}$, the ratio *r* and the positions of water molecules around $Na^+$ and in the interlayer space at large *r* values. $r^*$ is also shown.

It is observed a large difference in the frequency factors ($A$) in the two regimes of water desorption. Frequency factors could be associated with the number of attempts by molecule per unit time to escape from the matrix [26]. The comparison with desorption rates of water measured in starch grains [12] shows that the frequency factor is very small at the lowest moisture content (structural water), contrary to Lap. Ferrage et al. [27] report for Ca-Mt dehydration that the interlayer space is not homogeneous during the process, but do not stress upon any particular value of the moisture content.

## 4. Conclusions

This investigation highlights the differences between natural and synthetic (Lap RD) Sm type clays with respect to their hydration, swelling and desorption properties. Lap is a model system for natural clays. Crystalline swelling (interlayer distance) of Lap is fully controlled by the ratio *r*=$H_2O$/Na towards levels unreached with natural clays. Desorption kinetics in isothermal conditions allow identifying two regimes according to the hydration of the Na cations, at high (*r*>7) and low (*r*<7) water contents. The activation energies associated to the desorption processes in these regimes are 33kJ/mol and 52kJ/mol, respectively determined within our experimental conditions. Unexpectedly, the frequency factors are totally different in the two regimes. It is suggested that the mechanisms of water desorption of Sm type clays and in particular Lap, may constitute a special case and bring interesting new features about the unique role played by the cation stability.

### Acknowledgements
GAV gratefully acknowledges to CNPq (National Council for Scientific and Technological Development) for the ''SWE'' PhD (200247/2015-8) and São Paulo Research Foundation (FAPESP) Brazil, for the PhD fellowship (2012/24047-3) received between years 2013-2015. PJAS acknowledges to FAPESP for the research grant (CEPID, 2013/07914-8). The authors are very grateful to Dr. Sophie Nowak for her expertise in performing X-ray diffraction measurements.